\documentclass[aps,prb,twocolumn,nobibnotes]{revtex4-2}  
\usepackage{dcolumn}
\usepackage{graphicx}
\usepackage{color}
\usepackage{amssymb}
\usepackage{amsmath}
\usepackage{bm}
\usepackage{bbm}
\usepackage[colorlinks=true,linktocpage=true,breaklinks=true]{hyperref}
\usepackage{multirow}
\usepackage{braket}
\usepackage{appendix}
\usepackage{mathtools}

\begin{document}

\title{Orbital Kerr effect and terahertz detection via the nonlinear Hall effect}

\author{Diego Garc\'ia Ovalle}
\email{diego-fernando.garcia-ovalle@univ-amu.fr}
 \author{Armando Pezo}
 \email{armando-arquimedes.pezo-lopez@univ-amu.fr}
\author{Aur\'{e}lien Manchon}
\email{aurelien.manchon@univ-amu.fr}
\affiliation{Aix-Marseille Universit\'e, CNRS, CINaM, Marseille, France}

\date{\today}

\begin{abstract}
We investigate the optical response induced by a d.c. current flowing in a nonmagnetic material that lacks inversion symmetry. In this class of materials, the flowing current experiences a nonlinear Hall effect and induces a nonequilibrium orbital magnetization, even in the absence of spin-orbit coupling. As a result, an orbital-driven Kerr effect arises that can be used to probe not only the orbital magnetization but also the nonlinear Hall effect. In addition, in the long wavelength limit, the nonlinear Hall effect leads to a rectification current that can be used to detect terahertz radiation. We apply the theory to selected model systems, such as WTe$_2$ bilayer and metallic superlattices. The nonequilibrium orbital Kerr efficiencies obtained in these systems are comparable to the largest values reported experimentally in GaAs and MoS$_2$, exceeding the values reported in metals and suggesting a large terahertz current responsivity. 
\end{abstract}

\maketitle

\section{Introduction} 
\begin{figure}[ht!]
	\includegraphics[width=1\linewidth]{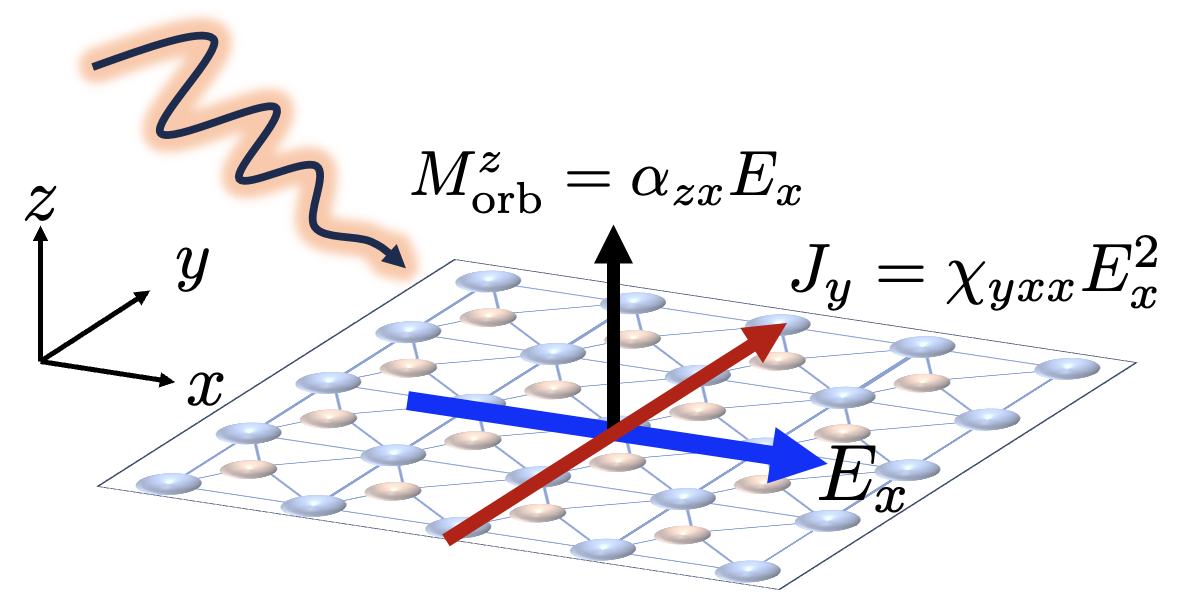}
    \caption{(Color online) Schematics of the connection between the nonequilibrium orbital Kerr effect, the orbital Edelstein effect ($\sim\alpha_{zx}$) and the nonlinear anomalous Hall effect ($\sim\chi_{yxx}$) in a noncentrosymmetric, nonmagnetic metal.}
	\label{Fig1}
\end{figure}

Nonreciprocal transport in noncentrosymmetric quantum materials has been attracting increasing interest over the past decade \cite{Tokura2018}, emphasizing the connection between nonlinear optical and electrical responses and the geometry of the Bloch states \cite{Ideue2017,Itahashi2020,Li2021}. Among these phenomena, the nonlinear Hall effect associated with the Berry curvature dipole \cite{Sodemann2015} (BCD) is particularly intriguing as it enables the generation of transverse currents even in the absence of magnetic field \cite{Ma2019,Kang2019}. In a recent proposal, Zhang and Fu \cite{Zhang2021b} suggested that the rectification induced by the nonlinear Hall effect could be used to detect terahertz radiations, a topic gaining increasing momentum \cite{Dhillon2017}. As discussed in detail below, the nonlinear Hall effect, at the second order in the electric field, is a companion of the nonequilibrium orbital magnetization \cite{Son2019}. Indeed, both effects only require inversion symmetry breaking and can emerge in the absence of spin-orbit coupling. Therefore, the nonlinear Hall effect is not only a signature of the BCD but also a signature of the nonequilibrium orbital magnetization. This connection is remarkably instrumental as it enables the optical probe of the nonequilibrium orbital moment via the so-called orbital-driven Kerr effect, as illustrated in Fig. \ref{Fig1}.

In recent years, it has been proposed that orbital currents and densities can be electrically induced in noncentrosymmetric metals via the so-called orbital Hall effect \cite{Bernevig2005b,Tanaka2008,Jo2018} and orbital Rashba-Edelstein effect \cite{Go2017,Yoda2018}, respectively, not necessitating spin-orbit coupling. This prediction opens wide perspectives for materials research and device development as entire families of metallic compounds made of light elements, often cheap and abundant, could in principle host interconversion phenomena between charge and orbital currents \cite{Jo2018,Salemi2021,Pezo2022,Salemi2022}. The central role of orbital currents in spin torque \cite{Ding2020,Lee2021b,Sala2022,Hayashi2023,Fukunaga2023}, spin pumping \cite{Hamdi2023} and magnetoresistance experiments \cite{Ding2022} has been clearly demonstrated recently.  

In all the experiments mentioned above, the clear distinction between spin and orbital transport phenomena remains challenging as both classes of effects are intermingled via spin-orbit coupling. Since the orbital accumulation of conduction electrons produces a nonequilibrium orbital magnetization, it can be detected by converting this orbital signal into a chemical potential using a proximate magnetic layer for instance. This technique is typically used to detect spin accumulation in magnetic bilayers \cite{Nakayama2013} or in nonlocal geometries \cite{Valenzuela2006}. When it comes to detecting the orbital accumulation though, a difficulty arises. Indeed, assuming that this proximate ferromagnet is made out of transition metals, its magnetization mostly comes from the spin angular momentum because of orbital quenching \cite{Blugel2007,Hanke2016}. Therefore, one needs to first convert the orbital signal into a spin signal, typically via spin-orbit coupling, as achieved in Ref. \cite{Ding2022}. In other words, the electrical detection of nonequilibrium orbital magnetization using a bilayer configuration requires the coexistence of spin-orbit coupling and magnetism. As a result, it is virtually impossible to selectively probe the orbital accumulation using electrical means only.  

An alternative method consists of using optical techniques \cite{Son2019,Choi2023,Lyalin2023} to directly probe the orbital accumulation without relying on any magnetic material or non-locality. Indeed, the magnetooptical Kerr (and Faraday) effect is routinely used to probe the magnetization of ferro- and ferrimagnets and image magnetic domains \cite{McCord2015}. In this process, the spin-orbit coupling enables the generation of the spin accumulation itself as well as its coupling with the circularly polarized light. This method has been successfully used in the past to detect the current-driven spin accumulation generated via spin Hall effect in GaAs \cite{Kato2004d}, InAs \cite{Lee2021c}, Pt and W thin films \cite{Stamm2017} and Bi$_2$O$_3$ surface \cite{Puebla2017}. In light metals though, both the generation of orbital accumulation out of equilibrium and its coupling with light occur in the absence of spin-orbit coupling, which results, in principle, in a much larger nonequilibrium Kerr signal \cite{Choi2023,Lyalin2023}.

In this Article, we show how the Kerr effect can be used to probe this nonequilibrium orbital magnetization, via the nonlinear anomalous Hall effect \cite{Sodemann2015}. In ferromagnets, the magnetooptical Kerr effect does not directly probe the magnetization itself, but it rather probes the anomalous Hall effect of the magnetic material, so that the Kerr angle is directly related to the anomalous Hall conductivity $\theta_k+i\phi_k\sim \sigma_{xy}/\sigma_{xx}$. Hence, the mere presence of the nonlinear Hall effect guarantees the current-induced orbital-driven Kerr effect. In addition, the rectified Hall current can be used to detect the presence of low-energy photons \cite{Zhang2021b}, typically in the terahertz range 1 THz $\approx$ 4 meV). Remarkably, the "current responsivity", i.e., the performance criterion for terahertz detection, is directly proportional to the orbital Kerr efficiency in the low-frequency limit, $R\propto\theta_k$. This relation opens appealing perspectives in the search for optimal materials for terahertz detection.

Our manuscript is organized as follows: In Section \ref{II} we present the general theory of the nonequilibrium orbital Kerr effect in nonmagnetic, noncentrosymmetric crystals, and establish the connection between the nonequilibrium orbital magnetization and the nonlinear Hall effect. Then, in Section \ref{III} we demonstrate the proof-of-principle on bilayer WTe$_2$ \cite{Du2018} and Nb$_{2n+1}$Si$_n$Te$_{4n+2}$ \cite{Zhao2023}. In Section \ref{IV}, we present first-principles calculations on a multilayer made out of transition metals. Finally, in Section \ref{V} we provide a general discussion and draw the main conclusions. 

\section{General Theory} \label{II}

\subsection{Relation between nonequilibrium orbital magnetization and nonlinear Hall effect}\label{AII}
Before addressing the nonequilibrium orbital Kerr effect, let us first clarify the relationship between nonequilibrium orbital magnetization and the nonlinear Hall effect. Since the seminal work of Thonhauser et al. \cite{Thonhauser2005} and Shi et al. \cite{Shi2007}, it is well established that the anomalous Hall effect and equilibrium orbital magnetization are both rooted in the Berry curvature of materials with broken time-reversal symmetry. Concretely, the anomalous Hall effect and the equilibrium orbital magnetization read
\begin{eqnarray}
    \sigma_{ij}&=&-\epsilon_{ijk}\frac{e^2}{\hbar}\int \frac{d^3{\bf k}}{(2\pi)^3}\sum_{n}\Omega_{n\mathbf{k}}^{k}f_{n\mathbf{k}}^{(0)},\label{def:ahe}\\
    {\bf M}_{\rm orb}&=&\int\frac{d^3{\bf k}}{(2\pi)^3}\sum_{n}\left({\bf m}_{n{\bf k}}+\frac{e}{\hbar}(\varepsilon_F-\varepsilon_{n{\bf k}}){\bm \Omega}_{n\mathbf{k}}\right)f_{n\mathbf{k}}^{(0)}.\nonumber\\\label{def:om}
\end{eqnarray}
Here, ${\bm \Omega}_{n\mathbf{k}}$ and ${\bf m}_{n{\bf k}}$ are the Berry curvature and orbital moment of band $n$ associated with Bloch state $|u_{n{\bf k}}\rangle$ \cite{Xiao2010b}. Besides, $f_{n\mathbf{k}}^{(0)}$ is the equilibrium Fermi distribution function. From now on, we will consider the Fermi distribution in the zero temperature limit., 
\begin{eqnarray}\label{eq:bc}
    {\bm \Omega}_{n\mathbf{k}}&=&-{\rm Im}\langle\partial_{\bf k}u_{n{\bf k}}|\times|\partial_{\bf k}u_{n{\bf k}}\rangle\\
    {\bf m}_{n{\bf k}}&=&\frac{e}{2\hbar}{\rm Im}\langle\partial_{\bf k}u_{n{\bf k}}|\times({\cal H}_{\bf k}-\varepsilon_{n{\bf k}})|\partial_{\bf k}u_{n{\bf k}}\rangle.\label{eq:om}
\end{eqnarray}
Clearly, the Berry curvature and orbital moment are collinear to each other, fulfill the same symmetry requirements, and their averages over the Fermi sea vanish when time-reversal symmetry is preserved. Remarkably though, both orbital magnetization and Hall conductivity become finite out-of-equilibrium in broken inversion symmetry systems. 

Let us now expand Eqs. \eqref{def:ahe} and \eqref{def:om} to the next order in the electric field, $\mathbf{E}$. Following the rationale developed by Sodemann and Fu \cite{Sodemann2015} for the nonlinear Hall effect and Yoda and Murakami \cite{Yoda2018} for the nonequilibrium orbital magnetization, the nonequilibrium Fermi distribution function reads $f_{n\mathbf{k}}=f_{n\mathbf{k}}^{(0)}-f_{n\mathbf{k}}^{(1)}$, and within the relaxation time approximation,
\begin{align}
    f_{n\mathbf{k}}^{(1)}=-e\tau(\mathbf{E}\cdot {\bf v}_{n{\bf k}}) \partial_{\varepsilon_{n\mathbf{k}}}f_{n\mathbf{k}}^{(0)},\label{06}
\end{align}
\noindent 
where $\tau$ is a constant scattering time and ${\bf v}_{n{\bf k}}=\partial_{\hbar{\bf k}}\varepsilon_{n\mathbf{k}}$. Replacing Eq. \eqref{06} into Eqs. \eqref{def:ahe} and \eqref{def:om}, we deduce that \cite{Konig2019,Bhowal2023}

\begin{eqnarray}
&&\sigma_{ij}=\epsilon_{ijk}\frac{e^3\tau}{\hbar}\int \frac{d^3{\bf k}}{(2\pi)^3}\sum_{n}\Omega_{n\mathbf{k}}^{k}(\mathbf{E}\cdot{\bf v}_{n{\bf k}})\partial_{\varepsilon_{n\mathbf{k}}} f_{n\mathbf{k}}^{(0)},\label{def:ahe2}\\
&&{\bf M}_{\rm orb}=-e\tau\int\frac{d^3{\bf k}}{(2\pi)^3}\sum_{n}{\bf m}_{n{\bf k}}(\mathbf{E}\cdot{\bf v}_{n{\bf k}})\partial_{\varepsilon_{n\mathbf{k}}} f_{n\mathbf{k}}^{(0)}.\label{def:om2}
\end{eqnarray}
In other words, a Hall current develops at the second order in the electric field \cite{Sodemann2015}, whereas a nonequilibrium orbital magnetization emerges at the first order \cite{Yoda2018},
\begin{eqnarray}
j_a&=&\frac{e^3\tau}{\hbar^2}\chi_{abc}E_bE_c,\label{eq:nlhe}\\
M_k&=&\frac{e^2\tau}{2\hbar^2}\alpha_{kb}E_b,\label{eq:neqom}
\end{eqnarray}
where 
\begin{eqnarray}
\chi_{abc}&=&\hbar\epsilon_{abk}\int \frac{d^3{\bf k}}{(2\pi)^3}\sum_{n}\Omega_{n\mathbf{k}}^{k}v^c_{n{\bf k}}\partial_{\varepsilon_{n\mathbf{k}}} f_{n\mathbf{k}}^{(0)},\label{eq:bcd}\\
\alpha_{kb}&=&-\frac{2\hbar^2}{e}\int\frac{d^3{\bf k}}{(2\pi)^3}\sum_{n}m^k_{n{\bf k}}v^b_{n{\bf k}}\partial_{\varepsilon_{n\mathbf{k}}} f_{n\mathbf{k}}^{(0)}.\label{eq:ore}
\end{eqnarray}
Equation \eqref{eq:bcd} is the Berry curvature dipole (BCD) \cite{Sodemann2015} and Eq. \eqref{eq:ore} is the orbital Edelstein effect \cite{Yoda2018}. In Eq. \eqref{eq:bcd}, one sees that the directions of the electric field $(b=c)$, Berry curvature ($k$), and Hall current ($a$) are orthogonal to each other. A symmetry analysis shows that the intrinsic nonlinear Hall effect is finite in the (x,y) plane only in $C_1$, $C_{1v}$, and $C_{2v}$ crystals, whereas extrinsic (skew and side-jump) nonlinear Hall effect (not addressed in the present work) also exists in trigonal ($C_3$, $C_{3v}$, $D_3$) and hexagonal crystals ($C_{3h}$, $D_{3h}$)  \cite{Sodemann2015,Du2021}. In contrast, the orbital Edelstein effect is present in any crystal lacking inversion symmetry \cite{Zelezny2017}, and it possesses in general both diagonal ($k=b$) and off-diagonal elements ($k\neq b$). Because the Berry curvature and orbital moment are collinear [see Eqs. \eqref{eq:bc}-\eqref{eq:om}], the nonlinear Hall effect is associated with the off-diagonal component of the orbital Edelstein effect. In other words, the nonlinear Hall effect and (off-diagonal) orbital Edelstein effects are companion phenomena. In the next section, we explain how magnetooptics can be used to detect the nonequilibrium orbital magnetization via the nonlinear Hall effect.

\subsection{Current-driven magnetooptics}\label{BII}

Without loss of generality, let us consider the magnetooptical Kerr effect in its polar version as depicted in Fig. \ref{Fig1} \cite{McCord2015,Oppeneer1999}. In this case, the electric permittivity tensor of the slab to be probed has the following form \cite{Oppeneer1999}
\begin{align}
    \hat{\varepsilon}=\begin{pmatrix}
\varepsilon_{xx} & \varepsilon_{xy}\\
\varepsilon_{yx} & \varepsilon_{xx}
\end{pmatrix}.\label{1}
\end{align}
\noindent 
Under the application of a perpendicularly incident light $\mathbf{{E}}_{\omega}=\mathcal{E} e^{i(\mathbf{k}\cdot\mathbf{z}-\omega t)}$ with frequency $\omega$, and assuming that $\varepsilon_{xx}\gg \varepsilon_{xy},\varepsilon_{yx}$, the Kerr rotation angle $\theta_k$ and the ellipticity $\phi_k$ can be approximated by \cite{Oppeneer1999},
\begin{align}
    \theta_k+i\phi_k=\frac{i\sqrt{\epsilon_{xy}\epsilon_{yx}}}{\sqrt{\epsilon_{xx}}(1-\epsilon_{xx})}.\label{1b}
\end{align}
\noindent 
Recalling the fundamental relation between the permittivity tensor and the conductivity tensor, $\hat{\varepsilon}=\hat{1}+\frac{4\pi i}{\omega} \hat{\sigma}(\omega)$, one obtains 
\begin{eqnarray}\label{2}
\theta_k+i\phi_k&=&-\frac{i\sqrt{\sigma_{xy}\sigma_{yx}}}{\sigma_{xx}\sqrt{1+\frac{4\pi i}{\omega}\sigma_{xx}}}.
\end{eqnarray}

The longitudinal conductivity in Eq. \eqref{2} can be accounted for through the Kubo-Streda formula \cite{Varga2020}
\begin{align}
    \sigma_{xx}=-\frac{e^2\hbar}{4\pi}\int \frac{d^2{\bf k}}{(2\pi)^2}\operatorname{Re}\left\{\operatorname{Tr}\left[\hat{v}_{\bf k}^xG_{\bf k}^{\rm R-A}\hat{v}_{\bf k}^xG_{\bf k}^{\rm R-A}\right]\right\}_{\epsilon_F},\label{3}
\end{align}
\noindent 
where $\hat{v}_\mathbf{k}=\partial_{\hbar{\bf k}}{\cal H}_{\bf k}$ is the velocity operator, and $G^{\rm R-A}_{\bf k}=G^{\rm R}_{\bf k}-G^{\rm A}_{\bf k}$, with $G^{\rm R(A)}_{\bf k}=\left[(\epsilon+\hbar\omega)-\mathcal{H}_{\mathbf{k}}\pm i\Gamma\right]^{-1}$. $\Gamma=\hbar/2\tau$ is the disorder broadening assuming a constant relaxation time $\tau$. In the absence of extrinsic skew or side jump scattering, the transverse optical conductivity is given by Eq. \eqref{def:ahe}, replacing the d.c. Berry curvature by the a.c. one,
\begin{eqnarray}
     \mathbf{\Omega}_{n\mathbf{k}}(\omega)=i\hbar^2\sum_{m\not=n}\frac{\bra{u_{n\mathbf{k}}}\hat{v}_\mathbf{k}\ket{u_{m\mathbf{k}}}\times\bra{u_{m\mathbf{k}}}\hat{v}_\mathbf{k}\ket{u_{n\mathbf{k}}}}{(\epsilon_{n\mathbf{k}}-\epsilon_{m\mathbf{k}})^2-(\hbar\omega +i\Gamma)^2}. \label{5}
\end{eqnarray}
This expression was obtained from Eq. \eqref{def:ahe} and applying the definition $|\partial_{\bf k}u_{n{\bf k}}\rangle=\hbar\sum_{m\not=n}\frac{\langle u_{m{\bf k}}|\hat{v}_{\bf k}|u_{n{\bf k}}\rangle}{\varepsilon_{n{\bf k}}-\varepsilon_{m{\bf k}}}|u_{m{\bf k}}\rangle$. From Eq. \eqref{5}, it is clear that if the system is time-reversal symmetric then $\sigma_{xy}\to 0$, and thus the magnetooptical Kerr effect vanishes in the absence of flowing current. Notwithstanding, when time reversal is preserved and inversion symmetry is broken, a nonequilibrium Kerr effect is allowed by symmetry, revealing the underlying generation of a nonequilibrium magnetization in the process. 

Since the magnetooptical Kerr effect is usually used to probe the magnetism at surfaces, it is convenient to express the nonlinear Hall effect, Eq. \eqref{eq:nlhe}, in two dimensions, i.e., choosing $k=z$. In this case, the nonlinear Hall current reads \cite{Sodemann2015} 

\begin{eqnarray}
{\bf j}=\frac{e^3\tau}{\hbar^2}({\bf D}\cdot{\bf E})({\bf z}\times{\bf E}),
\end{eqnarray}
where 
\begin{eqnarray}
{\bf D}=\int\frac{d^2{\bf k}}{(2\pi)^2}\sum_n \hbar{\bf v}_{n{\bf k}}\Omega^z_{n{\bf k}}\partial_{\epsilon_{n{\bf k}}}f_{n{\bf k}},\label{eq:bcd2d}
\end{eqnarray}
is the BCD in two dimensions. When a beam of light impinges on the slab while a d.c. electric field is applied, the total electric field is ${\bf E}_\omega+{\bf E}_0$, and the induced current reads, at the first order in ${\bf E}_\omega$, 

\begin{eqnarray}
{\bf j}=\frac{e^3\tau}{\hbar^2}\left[({\bf D}_0\cdot{\bf E}_0)({\bf z}\times{\bf E}_\omega)+({\bf D}_\omega\cdot{\bf E}_\omega)({\bf z}\times{\bf E}_0)\right].
\end{eqnarray}
The optical conductivity tensor $\hat{\sigma}(\omega)$, defined as ${\bf j}=\hat{\sigma}(\omega){\bf E}_\omega$, is therefore,
\begin{eqnarray}
\hat{\sigma}(\omega)=\begin{pmatrix}
-D_\omega^xE_0^y & -{\bf D}_0\cdot{\bf E}_0-D_\omega^yE_0^y\\
{\bf D}_0\cdot{\bf E}_0+D_\omega^x E_0^x & D_\omega^yE_0^x
\end{pmatrix}.
\end{eqnarray}
The optical Kerr effect is related to the antisymmetric part of the conductivity tensor $\hat{\sigma}^A(\omega)=(\hat{\sigma}(\omega)-\hat{\sigma}^T(\omega))/2$, which explicitly reads
\begin{eqnarray}
\hat{\sigma}_\omega^A=\begin{pmatrix}
0 & -({\bf D}_0+{\bf D}_\omega)\cdot{\bf E}_0\\
({\bf D}_0+{\bf D}_\omega)\cdot{\bf E}_0 & 0
\end{pmatrix}.
\end{eqnarray}
Finally, since ${\bf D}_0\gg{\bf D}_\omega=\frac{{\bf D}_0}{2(1+i\omega \tau)}$ for $\omega \tau\gg1$, we arrive at the non-equilibrium Kerr angle,
\begin{eqnarray}\label{neqKerr}
\theta_k+i\phi_k&\approx&\frac{{\bf D}_0\cdot{\bf E}_0}{\sigma_{xx}\sqrt{1+\frac{4\pi i}{\omega}\sigma_{xx}}}.
\end{eqnarray}
Therefore, the nonequilibrium Kerr effect is proportional to the BCD \cite{Sodemann2015}, which is the leading order contribution to the second-order optical conductivity at small frequencies \cite{Morimoto2016b}. As such, the nonequilibrium orbital magnetization, nonlinear Hall current, and nonequilibrium Kerr effect are companion phenomena in noncentrosymmetric nonmagnetic materials and heterostructures. Because of this intimate relationship between the orbital moment and the Kerr effect, we tag the latter "orbital Kerr effect". As complementary evidence of this connection, in the next section, we demonstrate a general decomposition of the nonequilibrium orbital magnetization in terms of the intrinsic BCD in the d.c. limit $\hbar\omega \to 0$. We stress out that although our discussion focuses on the intrinsic nonlinear Hall effect, Eq. \eqref{neqKerr} remains qualitatively valid for nonlinear Hall transport arising from extrinsic skew and side-jump scattering. This means that the connection between nonlinear Hall effect, orbital Edelstein effect, and nonequilibrium orbital Kerr effect is not limited to $C_1$, $C_{1v}$ and $C_{2v}$ crystals but also applies to $C_3$, $C_{3v}$, $D_3$, $C_{3h}$, and $D_{3h}$ systems \cite{Sodemann2015,Du2021}. Understanding the microscopic connection between these three quantities in the presence of extrinsic scattering lies out of the scope of the present study but is undoubtedly pertinent to interpreting experiments.

Before closing this subsection, let us comment on the rectification effect associated with the nonlinear Hall response. As pointed out in Ref. \cite{Zhang2021b}, a high-frequency electric field ${\bf E}_{\omega}= {\bf E}_0\cos(\omega t)$ induces an oscillatory nonlinear Hall current \cite{Sodemann2015}
\begin{eqnarray}
j_a=\frac{e^3\tau}{\hbar^2} \chi_{abc} E_bE_c \cos^2(\omega t).\label{resp}
\end{eqnarray}
The d.c. rectified Hall current $j^r_i=\langle j_i(t)\rangle$ averaged over a cycle period can be used to detect the incoming radiation. Following Ref. \cite{Zhang2021b}, the performance indicator of this detector is called the current responsivity $R$ and is given by the ratio between the Hall current $I_H$ and the absorbed power $P$,
\begin{eqnarray}
Rw=\frac{I_H}{P}=\frac{j^r}{\sigma_{xx}E_0^2}=\frac{\theta_k}{2E_0},
\end{eqnarray}
where $w$ is the width of the rectifier. Remarkably, the current responsitivity is simply equal to half of the orbital Kerr electrical efficiency.

\subsection{Decomposition of the nonequilibrium orbital magnetization} \label{CII}

Before closing this section, let us further comment on the connection between the nonlinear Hall effect and the nonequilibrium orbital magnetization. Explicitly, in the limit $\hbar\omega\to 0$, the orbital moment of band $n$, given in Eq. \eqref{eq:om}, reads \cite{Shi2007}
\begin{align}
    \mathbf{m}_{n\mathbf{k}}=\frac{ie\hbar}{2}\sum_{m\not=n} \frac{\braket{u_{n\mathbf{k}}|\hat{v}_{\mathbf{k}}|u_{m\mathbf{k}}}\times \braket{u_{m\mathbf{k}}|\hat{v}_{\mathbf{k}}|u_{n\mathbf{k}}}}{\epsilon_{n\mathbf{k}}-\epsilon_{m\mathbf{k}}}.\label{8}
\end{align}
Injecting this expression into the definition of the orbital Edelstein coefficient, Eq. \eqref{eq:ore}, and using the expression of the Berry curvature, Eq. \eqref{5}, we obtain
\begin{align}
&\alpha_{ij}(\mu)
=i\hbar^3\int\frac{d^2{\bf k}}{(2\pi)^2}\sum_{m\not=n,n}(\epsilon_{n\textbf{k}}-\epsilon_{m\textbf{k}})v_{n\mathbf{k}}^{j}\nonumber\\
&\left(\frac{\bra{u_{n\mathbf{k}}}\hat{v}_\mathbf{k}\ket{u_{m\mathbf{k}}}\times\bra{u_{m\mathbf{k}}}\hat{v}_\mathbf{k}\ket{u_{n\mathbf{k}}}}{(\epsilon_{n\textbf{k}}-\epsilon_{m\textbf{k}})^2}\right)_i\delta(\epsilon_{n\mathbf{k}}-\mu)\nonumber\\
&=-\int\frac{d^2{\bf k}}{(2\pi)^2}\sum_n \left(\epsilon_{n\textbf{k}}{\Omega}_{{n}\textbf{k}}^i-b_{{n}\textbf{k}}^i\right)v_{n\mathbf{k}}^{j}\delta(\epsilon_{n\mathbf{k}}-\mu)\nonumber\\
&=\mu D_{ji}(\mu) + \mathcal{B}_{ji}(\mu).\label{10}
\end{align}

\noindent 
As we can see from Eq. \eqref{10}, the nonequilibrium orbital magnetization can be divided into two contributions: a term linear in the chemical potential $\mu$ and proportional to the BCD [Eq. \eqref{eq:bcd2d}], and a term that is nonlinear in $\mu$ arising from an effective magnetic field with the same symmetries as the Berry curvature and the orbital magnetic moment,
\begin{align}
\textbf{b}_{n\mathbf{k}}&=i\hbar^2\sum_{m\not=n} \epsilon_{m\mathbf{k}}\frac{\bra{u_{n\mathbf{k}}}\hat{v}_\mathbf{k}\ket{u_{m\mathbf{k}}}\times\bra{u_{m\mathbf{k}}}\hat{v}_\mathbf{k}\ket{u_{n\mathbf{k}}}}{(\epsilon_{n\mathbf{k}}-\epsilon_{m\mathbf{k}})^2}.\nonumber\\
\end{align}
Remarkably, expression \eqref{10} goes one step forward from previous studies relating Berry curvature and orbital magnetic moment in the vicinity of the gap \cite{Du2018} and for two-band systems with particle-hole symmetry \cite{Xiao2007}.

\section{Two dimensional nonmagnetic systems} \label{III}

In this section, we compute the Kerr angle and analyze the previous decomposition for the current-induced orbital magnetization in the cases of WTe$_2$ bilayer and Nb$_{2n+1}$Si$_n$Te$_{4n+2}$ monolayer. Both systems display at $C_{1v}$ symmetry that enables a finite intrinsic BCD \cite{Sodemann2015,Du2021}.


\subsection{WTe$_2$ bilayer}

In order to illustrate our theory, we start by computing $\theta_k$ and $\phi_k$ in the exemplary case of the WTe$_2$ bilayer \cite{Du2018}, in which strong signatures of second-order Hall effect has been reported experimentally \cite{Ma2019,Kang2019}. The minimal model is composed of four tilted Dirac Hamiltonians in the form \cite{Du2018}

\begin{align}
\mathcal{H}_{\mathbf{k}}&=  \begin{pmatrix}
\mathcal{H}_{\mathbf{k}}^{d1} & \mathcal{P}_{\mathbf{k}} & 0 & \gamma\\
\mathcal{P}_{\mathbf{k}}^{\dag} & \mathcal{H}_{\mathbf{k}}^{d1} & \gamma & 0\\
0 & \gamma & \mathcal{H}_{\mathbf{k}}^{d2} & \mathcal{P}_{\mathbf{k}}\\
\gamma & 0 & \mathcal{P}_{\mathbf{k}}^{\dag} & \mathcal{H}_{\mathbf{k}}^{d2}
\end{pmatrix},  \label{12}
\end{align}
\noindent 
where $\mathcal{H}_{\mathbf{k}}^{di}$ describes the tilted Dirac cone at $\mathbf{K_i}$,
\begin{align}
   \mathcal{H}_{\mathbf{k}}^{di}&= [E_i+t_i(k_x+K_i)]\hat{\sigma}_0\nonumber\\
   &+v_i[k_y\hat{\sigma}_1 +\eta_i(k_x+K_i)\hat{\sigma}_2] + \frac{m_i\hat{\sigma}_3}{2}, \label{13}
\end{align} 
\noindent 
where $\hat{\sigma}_i$, $i=1..3$ are the Pauli spin matrices, $\hat{\sigma}_0=\mathrm{1}_{2\times2}$, and $\gamma$ is an interlayer coupling between the layers. In addition, 
\begin{align}
 \mathcal{P}_{\mathbf{k}}&= \begin{pmatrix}
 \nu_xk_x-i\nu_yk_y & 0 \\
 0 & -\nu_xk_x-i\nu_yk_y
 \end{pmatrix}\label{14}
\end{align} 
\noindent 
is a matrix that contains the spin-orbit coupling strength $\nu_x$ ($\nu_y$) along $\mathbf{x}$ ($\mathbf{y}$). We emphasize that although none of the phenomena we investigate here - nonlinear Hall effect, orbital Edelstein effect, and orbital Kerr effect - necessitate spin-orbit coupling, they may be highly sensitive to it. This is particularly true in the present case, where spin-orbit coupling-driven Dirac states dominate transport properties. Since the Berry curvature of the eigenstates of Eq. (\ref{12}) is highly concentrated around the Dirac cones \cite{Du2018}, we perform our calculations in the domain $\left[-\pi,\pi\right]\times\left[-\pi,\pi\right]$ for simplicity. The Berry curvature and the nonequilibrium orbital moment are perpendicular to the $(k_x,k_y)$ plane, and since the system also has a mirror symmetry along $k_y$, only $D_{xz}$ and $\alpha_{zx}$ are non zero. Therefore, when the electric field is applied along $\hat{x}$ and the Berry curvature is along $\hat{z}$, the resulting orbital magnetization is along $\hat{z}$ and the Hall current flows along $\hat{y}$. From the symmetry perspective, we can interpret the second-order Hall current along $\hat{y}$ as the interplay between the nonequilibrium orbital magnetization along $\hat{z}$ and the external electric field applied along $\hat{x}$. 

\begin{figure}[ht!]
	\includegraphics[width=1\linewidth]{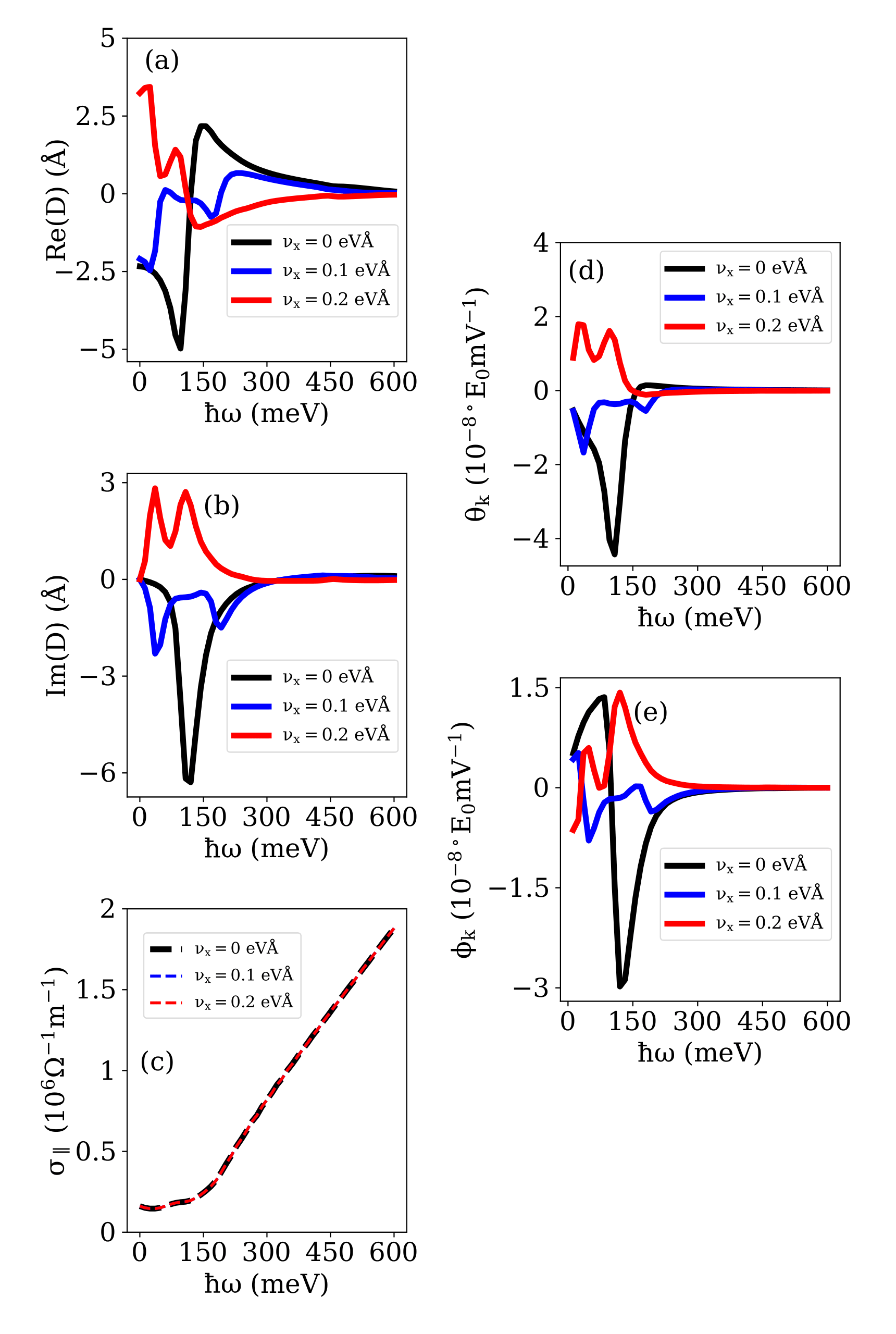}
    \caption{(Color online) Real (a) and imaginary (b) part of the a.c. BCD as a function of the energy of the incident light for different values of spin-orbit coupling $\nu_x$ and taking $\mu=0$. (c) Corresponding longitudinal conductivity, (d) Kerr rotation, and (e) ellipticity. We consider the values adopted in Ref. \cite{Du2018}: $\nu_y=0$, $\gamma=0.05$ eV, $K_1=0.1\pi$ \AA$^{-1}$, $K_2=0.15\pi$ \AA$^{-1}$, $v_1=v_2=2$ eV\AA, $t_1=t_2=1.5$ eV \AA, $m_1=m_2=0.1$ eV, $\eta_1=-\eta_2=-1$, $E_1=0.02$ eV, $E_2=-0.08$ eV and $\Gamma = 0.01$ eV.}
	\label{Fig2}
	\end{figure}
	
The real and imaginary parts of the a.c. BCD and the longitudinal conductivity as a function of the incoming light energy are reported in Figs. \ref{Fig2}(a,b) and (c), respectively. While the real part of the BCD converges to a finite but non-zero value at $\hbar\omega \to 0$, the imaginary part goes to zero at zero frequency. In addition, $\operatorname{Re}(D)$ and $\operatorname{Im}(D)$ are concentrated at energies $\hbar\omega \leq 250$ meV, corresponding to the model system's bandwidth. The longitudinal conductivity [Fig. \ref{Fig2}(c)] steadily increases with $\hbar\omega$ and remains mostly independent of the spin-orbit coupling parameter $\nu_x$. In contrast, both real and imaginary parts of the BCD are highly sensitive to spin-orbit coupling and present a sharply peaked structure. Nonetheless, there is no simple relationship between the BCD peaks and peculiarities in the band structure (avoided band crossing and Berry curvature maxima). From the BCD, one can compute the Kerr angle and ellipticity, displayed in Figs. \ref{Fig2}(d,e), that show a similar energy dependence as the BCD: at $\nu_x=0$, the peak around $125$ meV corresponds to the maximum of $\operatorname{Re}(D)$ and $\operatorname{Im}(D)$ for the same energy. For $\nu_x \not =0$, the average value of $\theta_k$ and $\phi_k$ becomes large by increasing the spin-orbit coupling. The high sensitivity of the nonlinear Hall effect and orbital Kerr response on spin-orbit coupling suggests that these properties can be controlled externally with a gate voltage \cite{Du2018}. 

The terahertz current responsivity, defined in Eq. \eqref{resp}, is readily obtained from Fig. \ref{Fig2}(d). Contrary to what was claimed in Ref. \cite{Zhang2021b}, the responsivity is frequency-dependent, which means that it can be optimized through band structure engineering. The current responsivity obtained at 50 meV ($\equiv$12 THz$\equiv$3.3 mm) is about $10^{-2}$ (A$\cdot\mu$m)/W, midway between GeTe and NbP \cite{Zhang2021b}.

\begin{figure}[ht!]
	\includegraphics[width=1\linewidth]{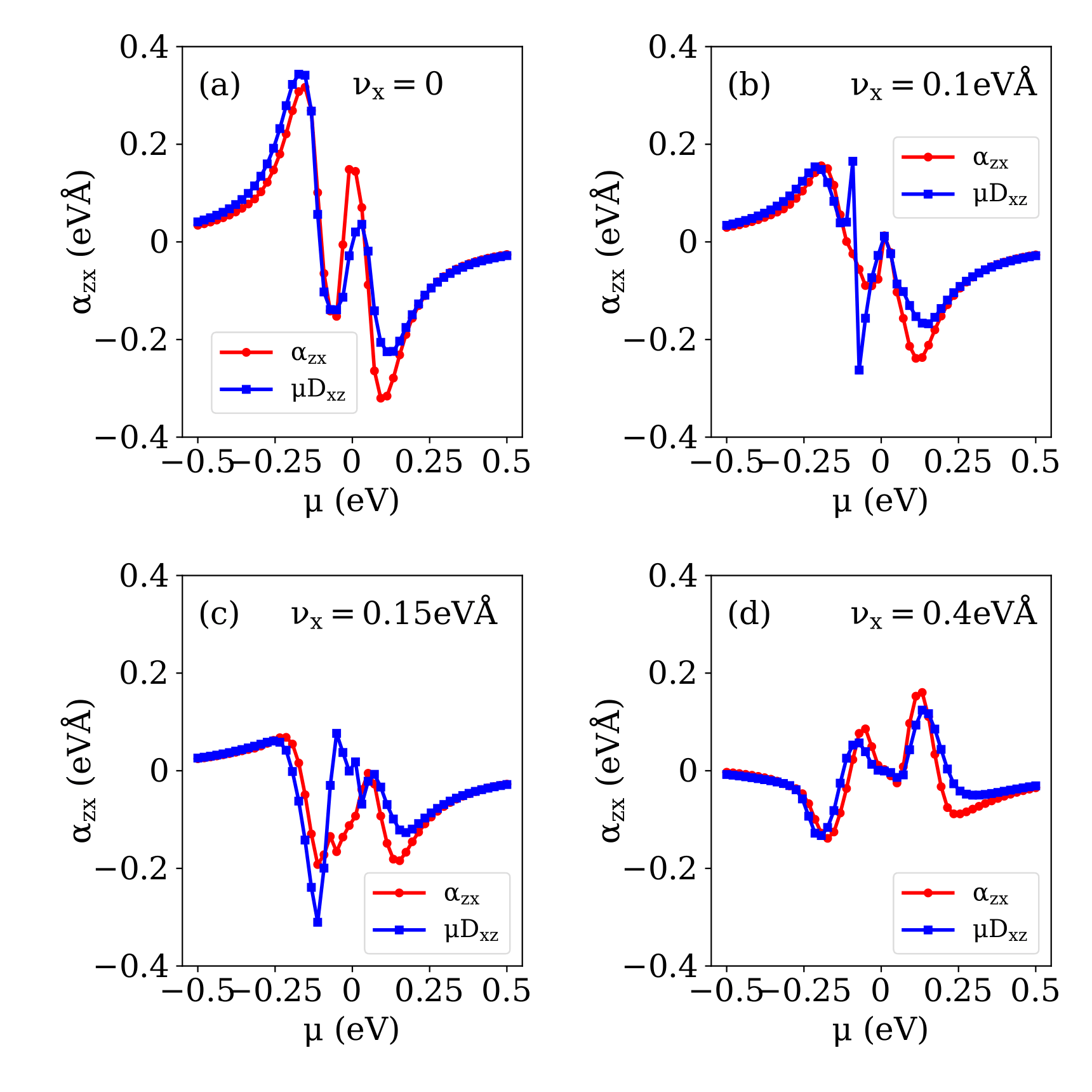}
	\caption{(Color online) Comparison between the orbital Edelstein coefficients and the BCD for the minimal model of WTe$_2$ bilayer, for different values of $\nu_x$ and as a function of $\mu$. We fix the same microscopic parameters that in Fig. \ref{Fig2}, exemplifying the cases (a) $\nu_x=0$, (b) $\nu_x=0.1$ eV \AA, (c) $\nu_x=0.15$ eV \AA and (d) $\nu_x=0.4$ eV \AA.}
	\label{Fig3}
\end{figure}

As an additional probe of the connection between the nonequilibrium orbital magnetization and the second-order Hall effect, in the d.c. limit $\hbar\omega\to 0$, we evaluate the decomposition of the orbital magnetization given by Eq. \eqref{10}. In fact, such a relationship has been proposed by Son et al. \cite{Son2019} in the vicinity of the gap of strained MoS$_2$, and we intend to assess the validity of this connection in the more complex case of WTe$_2$ bilayer. As explained above, the low-energy Hamiltonian of WTe$_2$ couples four tilted Dirac cones defined as $h_0=E_i+t_i(k_x+K_i)$. The term $E_i+t_iK_i$ results in a rigid energy shift and $t_ik_x$ is an odd function in $k_x$. Consequently, we expect that the first term in Eq. \eqref{10} must dominate over $\mathcal{B}$ close to the neutrality point. In other words, the orbital Edelstein effect in WTe$_2$ bilayer must follow $\alpha_{zx}\simeq \mu D_{zx}$, as long as the chemical potential $\mu$ remains in the vicinity of the Dirac nodes. These two terms, $\alpha_{zx}$ and $\mu D_{zx}$, are reported in Fig. \ref{Fig3} as a function of $\mu$ for different values of the spin-orbit coupling. When $\nu_x=0$ [Fig. \ref{Fig3}(a)], the system possesses two tilted Dirac cones and $\alpha_{zx}\simeq \mu D_{xz}$. Upon increasing the spin-orbit coupling [Figs. \ref{Fig3}(b,c)], the system exhibits band inversion and anticrossing (see Ref. \cite{Du2018}) such that the linear relationship between the nonlinear Hall effect and nonequilibrium orbital moment fails due to discontinuity in the Berry curvature related to topological transitions. At $\nu_x=0.4$ eV \AA [Fig. \ref{Fig3}(d)], a gap opens and the linear relation between $\alpha_{zx}$ and $D_{xz}$ is restored. In other words, the validity of the phenomenological relationship between the BCD and the orbital Edelstein effect claimed in Ref. \cite{Son2019} is not universal and only holds as long as the band structure can be approximated by a gapped Dirac cone (as in MoS$_2$ and numerous transition metal dichalcogenides).

\subsection{Nb$_{2n+1}$Si$_n$Te$_{4n+2}$ monolayer}

Large nonlinear Hall effect and orbital Kerr effect are expected to be found in materials displaying diverging Berry curvature close to Fermi level, i.e., typically Dirac or Weyl semimetals \cite{Armitage2018}. Recently, it has been proposed that the family of van der Waals compounds Nb$_{2n+1}$Si$_n$Te$_{4n+2}$ can also display Dirac nodes \cite{Yang2020b} and nodal lines \cite{Liu2023}. A model Hamiltonian for such a system reads
$\mathcal{H}_{\mathbf{k}}=\mathcal{H}_{\mathbf{k}}^{(0)}+\mathcal{H}_{\mathbf{k}}^{(soc)}$, where \cite{Zhao2023}

\begin{align}
   \mathcal{H}_{\mathbf{k}}^{(0)}&=t \begin{pmatrix}
  0 & 1+e^{-ik_x} \\
  1+e^{ik_x}  & 0
 \end{pmatrix}\otimes \hat{\sigma}_0  \nonumber\\
 &+\delta t\begin{pmatrix}
  0 & e^{-ik_y} (1+e^{-ik_x})\\
  e^{ik_y} (1+e^{ik_x}) & 0
 \end{pmatrix}\otimes \hat{\sigma}_0,\label{15}\\
   \mathcal{H}_{\mathbf{k}}^{(soc)}&=t \begin{pmatrix}
  \lambda_1\sin k_x & 0 \\
 0  & -2\lambda_1\sin k_x
 \end{pmatrix}\otimes \hat{\sigma}_3 + \nonumber\\
 &+\delta t\begin{pmatrix}
  2\lambda_3\sin k_y & i\lambda_2e^{ik_y} (1+e^{-ik_x})\\
  -i\lambda_2e^{-ik_y} (1+e^{ik_x}) & 2\lambda_3\sin k_y
 \end{pmatrix}\otimes \hat{\sigma}_3.\label{16}
\end{align}

\begin{figure}[ht!]
\includegraphics[width=\linewidth]{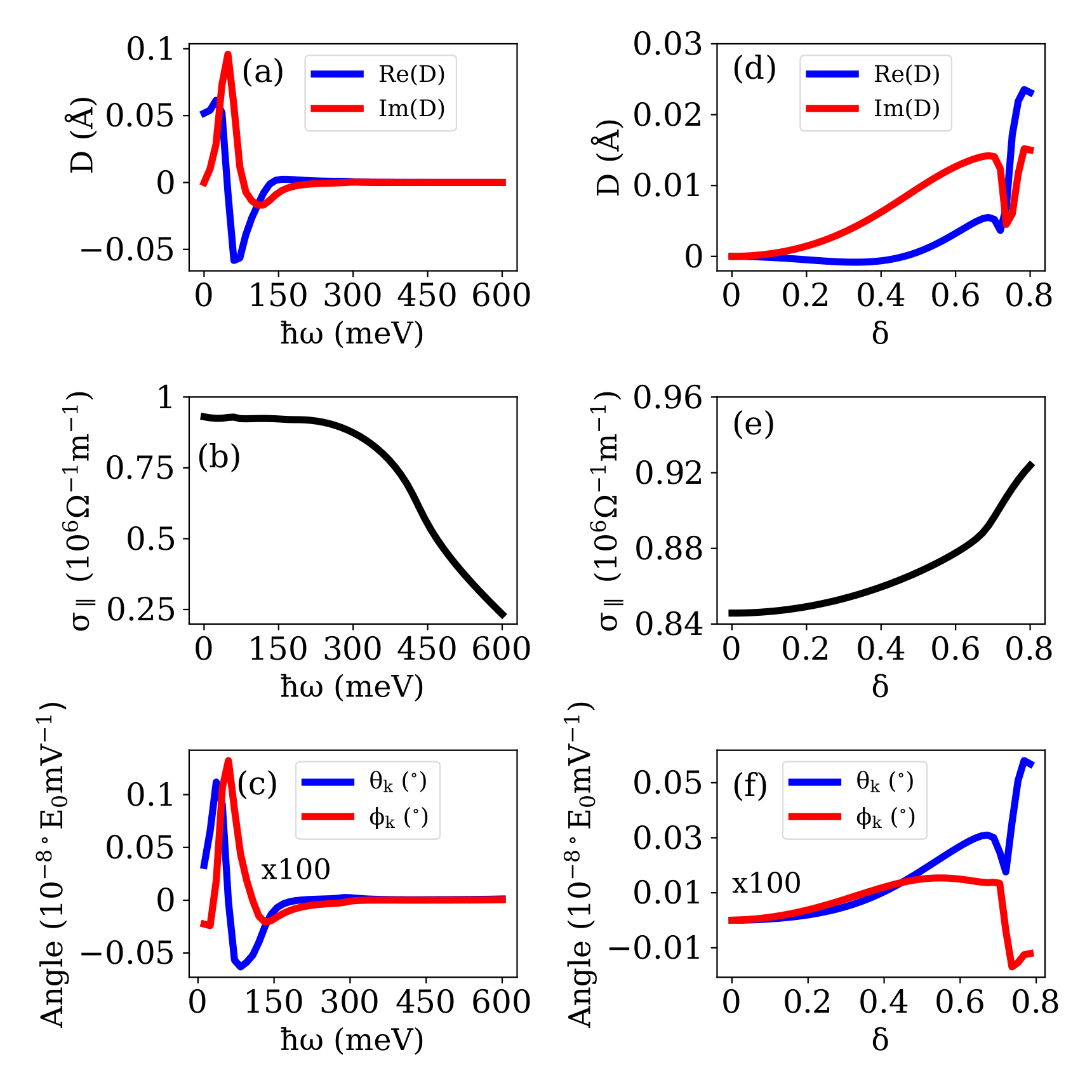}
\caption{(Color online) (a) Real (blue line) and imaginary (red line) part of the a.c. BCD as a function of the energy of the incident light for the Nb$_{2n+1}$Si$_n$Te$_{4n+2}$ monolayer. (b) Longitudinal conductivity of the system, (c) Kerr angle (blue line), and ellipticity (red line) as a function of the energy of the incident light. (d,e,f) Same quantities as a function of inter-chain parameter $\delta$. We set the parameters to $t=-0.2$ eV, $t'=\delta t= -0.16$ eV, $\lambda_1=1$, $\lambda_2=\lambda_3=0.1$, $\mu=-0.028$ eV, $\hbar\omega=100$ meV and $\Gamma=0.01$ eV.}
\label{Fig4}
\end{figure}

In Eqs. \eqref{15}-\eqref{16}, $t$ and $\delta t$ are the intra and inter-chain hopping parameters, respectively, and $\lambda_i$ ($i=1...3$) are the Rashba parameters. Considering that the Berry curvature and the orbital moment are parallel to $\hat{z}$, the nonequilibrium orbital magnetization is along $\hat{z}$ and a current flowing along $\hat{y}$ generates a second-order Hall current along $\hat{x}$ with the BCD coefficient $D_{yz}$. Let us first study the nonequilibrium orbital Kerr effect in this system, showing the behavior of the BCD, the longitudinal conductivity, and the Kerr angle as a function of the energy of the incident light. Our results are reported in Fig. \ref{Fig4}.

\begin{figure}[ht!]
\includegraphics[width=\linewidth]{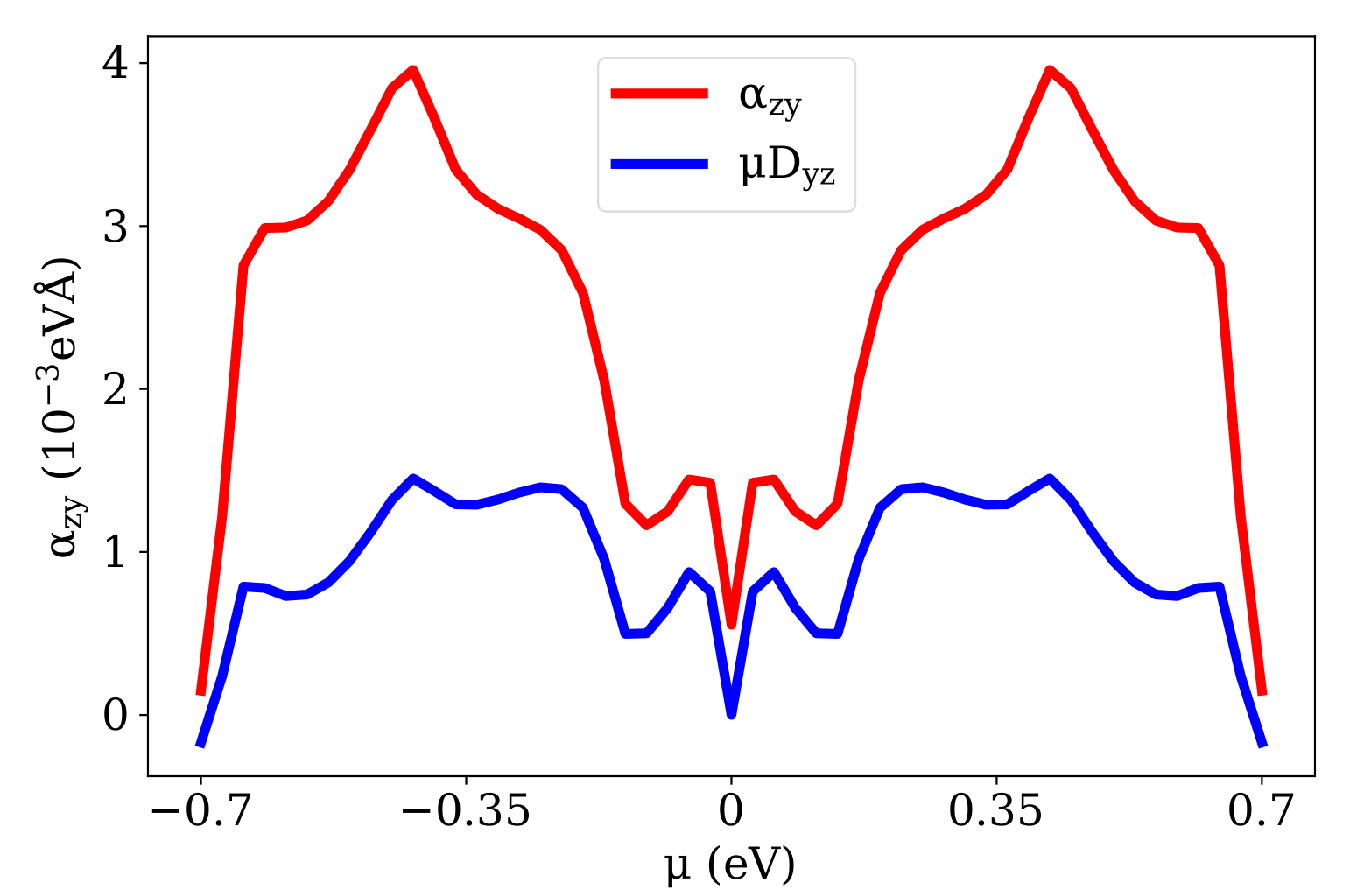}
\caption{(Color online) Orbital Edelstein coefficient $\alpha_{zy}$ and BCD coefficient $\mu D_{yz}$ as a function of the chemical potential for the Nb$_{2n+1}$Si$_n$Te$_{4n+2}$ monolayer. The microscopic parameters are $t=-0.2$ eV, $t'=\delta t= -0.16$ eV, $\lambda_1=1$, $\lambda_2=\lambda_3=0.1$ and $\mu=-0.028$ eV.}
\label{Fig5}
\end{figure}

In Fig. \ref{Fig4}(a), we verify that at zero frequency the real part of the BCD is finite while its imaginary part vanishes, as expected. The coefficients also tend to zero upon increasing the frequency, with peaks close to $\hbar\omega \to 100$ meV. Besides, the longitudinal conductivity is almost a constant for the region $\hbar\omega\leq 300$ meV [Fig. \ref{Fig4}(b)] and then decreases slowly with the frequency. This signal has the same order of magnitude as the response reported in previous experiments on Nb$_3$SiTe$_6$ \cite{Hu2015c,Allah2023}, whereas the nonequilibrium orbital Kerr effect is notoriously smaller than the one computed in WTe$_2$ because of the differences in the BCD and the longitudinal conductivity. We also find that restoring inversion symmetry by tuning the inter-chain hopping to zero progressively quenches the BCD [Fig. \ref{Fig4}(d,e)] and nonequilibrium orbital Kerr effect [Fig. \ref{Fig4}(f)]. Finally, Fig. \ref{Fig5} shows the orbital Edelstein coefficient and BCD, $\alpha_{zy}$ and $\mu D_{yz}$, as a function of the chemical potential $\mu$ in the d.c. limit. Whereas both functions have a similar behavior, their magnitude is markedly different. This again suggests that although both the nonequilibrium orbital moment and nonlinear anomalous Hall effect are companion phenomena that exist under the same symmetry-breaking conditions, an explicit relationship between these two mechanisms can only be established in the limiting case of gapped Dirac systems. 

\section{Orbital Kerr effect in metallic multilayers} \label{IV}

\begin{figure}[ht!]
\includegraphics[width=1\linewidth]{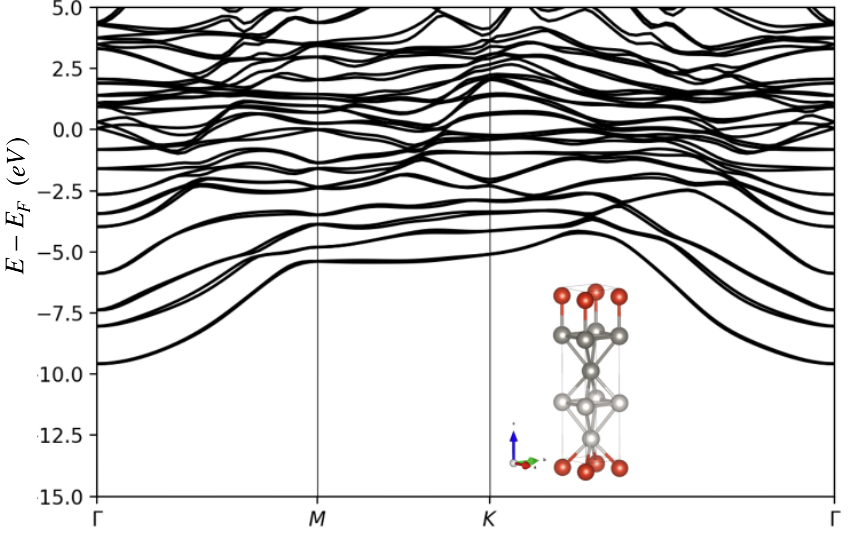}
\caption{(Color online) Electronic band structure of W(2)/Pt(2)/V(1) superlattice stacked along (110). The unit cell is depicted in the inset.}
\label{Fig6}
\end{figure}

To illustrate the emergence of the orbital Kerr effect in a realistic system, we perform first-principles simulations on a superlattice made of an infinite repetition of W(2)/Pt(2)/V(1) trilayers, a rather conventional system made of transition metals, familiar to mainstream experimental spintronics. We have used a plane-wave basis set to obtain the optimized structures employing the Perdew-Burke-Ernzerhof (PBE) \cite{gga,pbe} exchange-correlation functional implemented in the \textsc{VASP} package \cite{vasp1,vasp2}. In these simulations, we used 400 eV for the plane-wave expansion cutoff, and ionic potentials are described using the projector augmented-wave (PAW) method \cite{paw} performing the geometry optimization with a force criterion of 5$\times10^{-3}$ eV/\AA. The band structure is shown in Fig. \ref{Fig6}, where the Fermi level is set to zero and the unit cell structure is shown in the inset. The point group of the superlattice is C$_{2v}$ and, according Ref. \cite{Sodemann2015}, the only nonzero elements of the BCD tensor are $D_{xy}$ and $D_{yx}$.

\begin{figure}[ht!]
\includegraphics[width=1\linewidth]{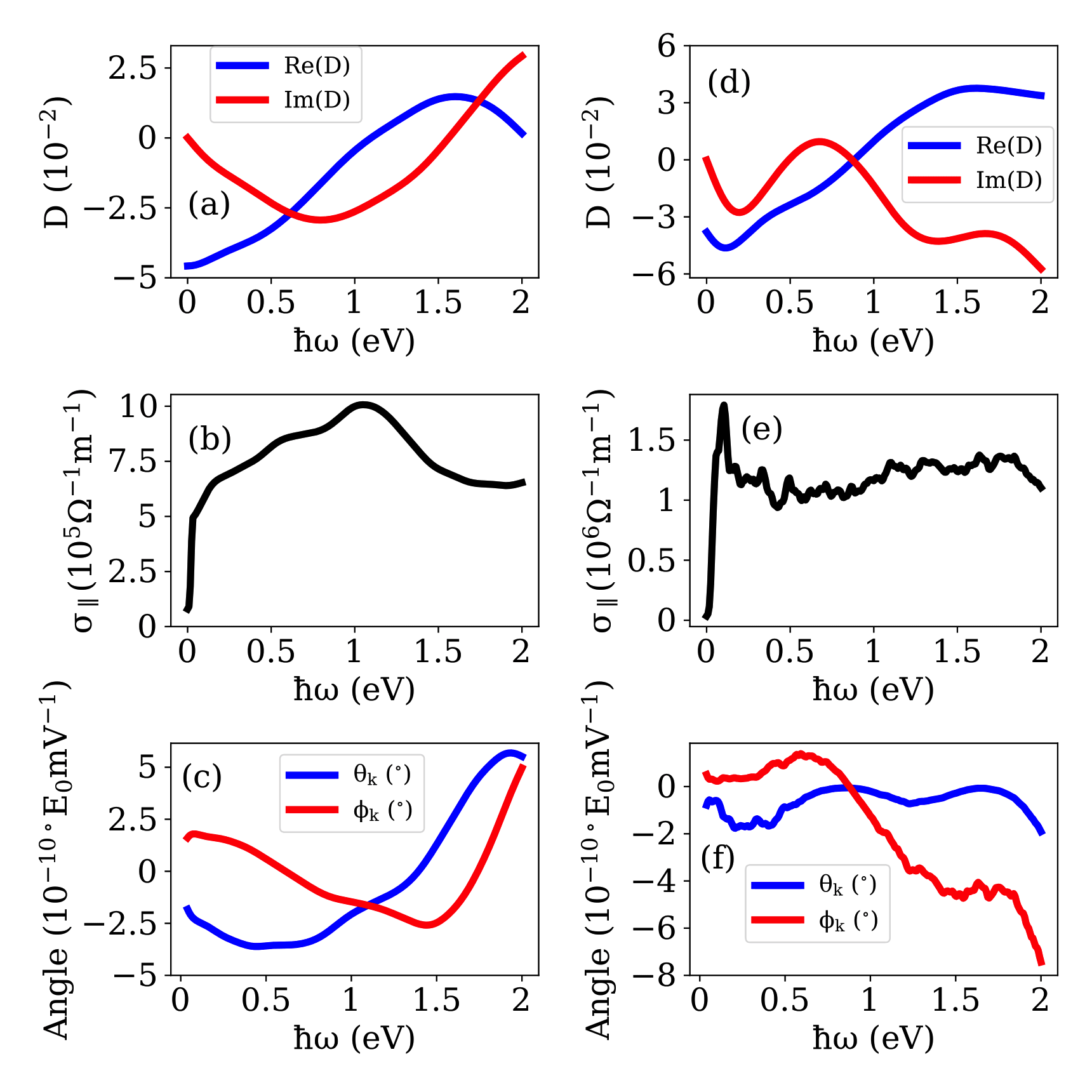}
\caption{(Color online) (a,d) Real (blue line) and imaginary (red line) parts of the BCD, (b,e) longitudinal conductivity and (c,f) nonequilibrium orbital Kerr effect for the W(2)/Pt(2)/V(1) superlattice, (a,b,c) without and (d,e,f) with spin-orbit coupling, as a function of the energy of the incident light.}
\label{Fig7}
\end{figure}

The BCD (a, d), the longitudinal conductivity (b, e), and the resulting nonequilibrium orbital Kerr effect (c, f) of the W(2)/Pt(2)/V(1) superlattice are reported in Fig. \ref{Fig7} without (left panels) and with spin-orbit coupling (right panels). Remarkably, the nonlinear Hall and nonequilibrium orbital Kerr effects, do not necessitate spin-orbit coupling. Although W and Pt possess a large spin-orbit coupling, the overall magnitude of these different mechanisms is about the same with or without spin-orbit coupling. We emphasize that the inversion symmetry is broken along $\hat{z}$, which means that the orbital Edelstein effect generates a nonequilibrium orbital moment in the plane perpendicular to $\hat{z}$. The electric field is injected along $\hat{x}$, producing a non-equilibrium orbital magnetization along $\hat{y}$ and a nonlinear Hall effect along $\hat{z}$. To measure the second-order Hall current experimentally in the W(2)/Pt(2)/V(1) superlattice, one needs to probe the current that flows {\em perpendicular} to the plane of the stack, which is a rather unusual configuration. 

\begin{figure}[ht!]
\includegraphics[width=1\linewidth]{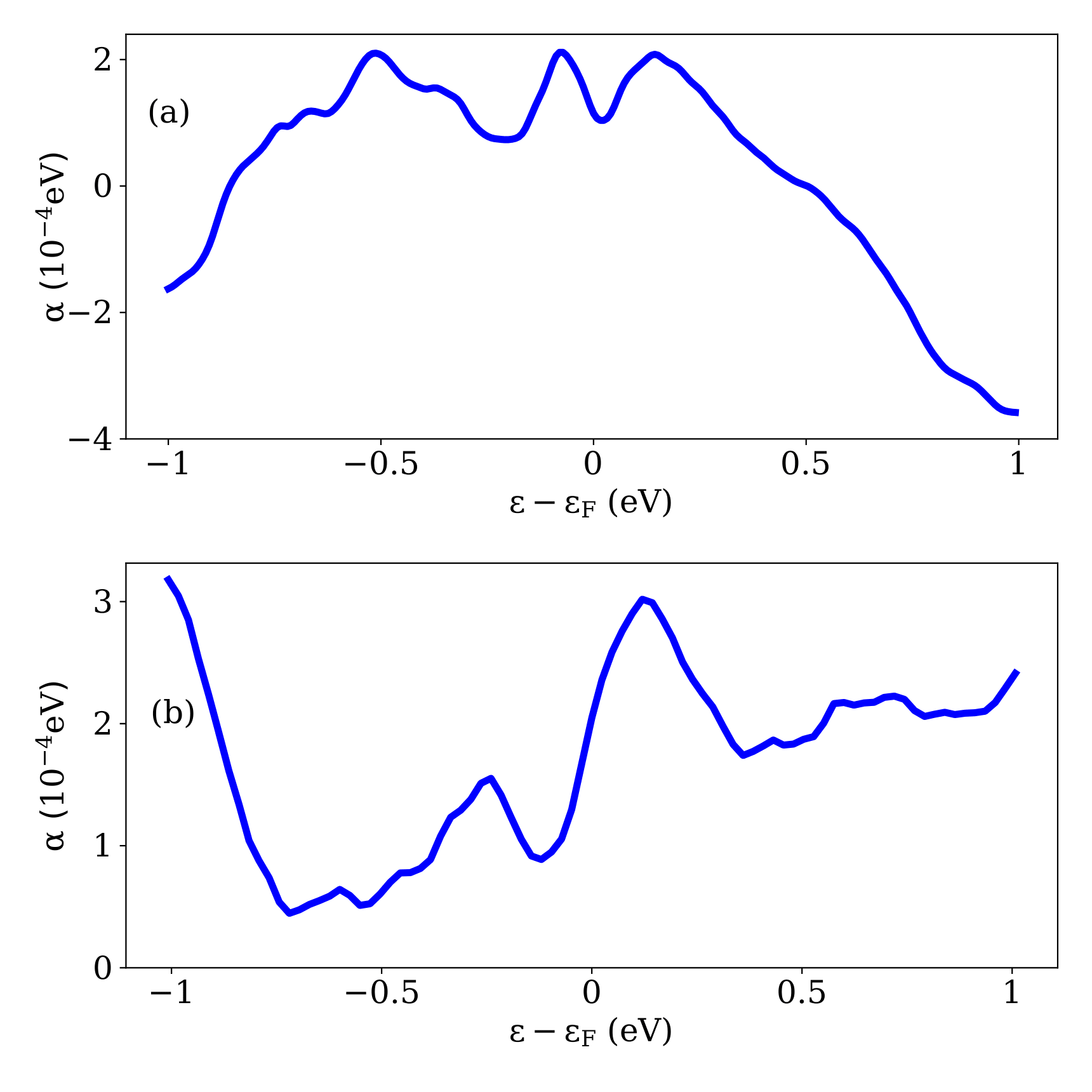}
\caption{(Color online) Orbital Edelstein coefficient as a function of the carrier energy for W(2)/Pt(2)/V(1) superlattice, (a) without and (b) with spin-orbit coupling.}
\label{Fig8}
\end{figure}

To complete this study, the orbital Edelstein coefficients of W(2)/Pt(2)/V(1) superlattice are displayed in Fig. \ref{Fig8} as a function of the carrier energy, (a) without and (b) with spin-orbit coupling. To compare with the value computed in WTe$_2$ bilayer, one needs to divide the orbital Rashba coefficient of WTe$_2$ bilayer by the thickness, giving an order of magnitude of about $10^{-3}$, about one order of magnitude larger than the orbital Edelstein coefficients computed in the W(2)/Pt(2)/V(1) superlattice. This result suggests that interfacial engineering could be a practical path to obtain large current-driven orbital magnetization and nonlinear Hall effect.

\section{Discussion and Conclusion}\label{V}

In order to assess the ability of the orbital Kerr effect to probe the nonequilibrium orbital accumulation in realistic situations, we now compare our results with experimental data obtained on different material platforms. We collected the value of the nonequilibrium orbital Kerr effect efficiencies, $\theta/j$ and $\theta/E_0$, for various systems published in the literature in Table \ref{TableI}, together with representative values obtained in the present work. We stress that comparing experiments is not an easy task because the conductivity of the materials (from $10^3$ to $10^{7}$ $\Omega^{-1}\cdot$m$^{-1}$) and the wavelength of light used for magnetooptical detection (from 10.6 $\mu$m to 514 nm) can be vastly different. 

\begin{table}[ht!]
\begin{tabular}{c|ccccc}
Material & $\lambda$ & $\sigma$ &  $\theta/j$ & $\theta/E_0$ & Ref. \\
 & nm & $\Omega^{-1}\cdot$m$^{-1}$ &  rad/(A/cm$^2$) & rad/(V/m) &  \\ \hline\hline
  WTe$_2$ &10.6$\times10^3$& 1.9$\times10^{5} $ & -3.1$\times10^{-11}$ & -5.8$\times10^{-10}$ & \\ 
  NbSiTe &$10.6\times10^3$& 9.2$\times10^{5} $ & -8$\times10^{-16}$ & -7.4$\times10^{-14}$ & \\ 
  W/Pt/V&$10.6\times10^3$& 1.6$\times10^{6} $ & -1.5$\times10^{-14}$ & -2.4$\times10^{-12}$ & \\ 
\hline
GaAs     &825& $2.8\times 10^{3}$& $7\times 10^{-10}$ & $2\times 10^{-10}$ & \cite{Kato2004d} \\ 
MoS$_2$ &$\sim$660& $10^3$ & $8.5 \times 10^{-11}$ & $8.5 \times 10^{-12}$ & \cite{Son2019} \\ 
Pt  & 514 & $6\times10^{6}$ &  $2.3\times 10^{-15}$ &$1.4\times 10^{-12}$ & \cite{Stamm2017} \\ 
Pt  &800& $7\times10^{6}$ &  $3\times10^{-16}$ & $2\times 10^{-13}$ & \cite{Lyalin2023} \\ 
W  &514 & $6\times 10^{5}$ &  -$6.6\times 10^{-15}$ & -$4\times10^{-13}$ & \cite{Stamm2017} \\ 
Ti &780& $2.5\times10^{6}$ & -$5\times10^{-15}$ & -$1.2\times 10^{-12}$ & \cite{Choi2023} \\ 
Cr  &800& $5\times10^{6}$& -$1.2\times10^{-15}$ & -$6\times 10^{-13}$ & \cite{Lyalin2023} \\ 
\end{tabular}
\caption{Summary of the magnetooptical Kerr efficiencies calculated (top) and measured (bottom) in different materials. (top) In the theoretical data, we fixed the wavelength of the incident light to $\hbar\omega=117$ meV. (bottom) The experimental values have been extracted from the literature when available.}\label{TableI}
\end{table}

Let us start by considering the results obtained on GaAs in the seminal work of Kato et al. \cite{Kato2004d}. Except for Te, the electrical orbital Kerr efficiency obtained by Kato et al. is the largest reported to far, $\theta_{\rm GaAs}/E_0 \simeq 2\times 10^{-10}$ rad m V$^{-1}$, which is comparable to our numerical results for the WTe$_2$ bilayer and the W/Pt/V superlattice. Interestingly, this observation is interpreted in the framework of the spin Hall effect, which is rather surprising considering the very small spin-orbit coupling of GaAs. Although it would require further investigation, we speculate that this unexpectedly large value could perhaps be related to the nonequilibrium orbital accumulation rather than to spin accumulation.

When considering the ratio between the Kerr angle and the current, $\theta/j$, we see a drastic difference between semiconductors (GaAs, and MoS$_2$), and metallic systems (Cr, Ti, Pt, and W). The formers display an orbital Kerr efficiency that is about four orders of magnitude larger than their metallic counterpart. When considering the ratio $\theta/E_0$, the orbital Kerr efficiency of MoS$_2$ remains one order of magnitude larger than in metals. The theoretical values computed in the present work are all larger than the values collected in metallic systems. 

The connection between the orbital Kerr effect and the current responsivity to terahertz radiation is particularly interesting. We found (semi)metallic systems like WTe$_2$ and W/Pt/V superlattices, that possess a relatively large BCD, display a smaller responsivity due to their high conductivity ($Rw\approx 10^{-4}$ (A$\cdot\mu$m)/W). This finding opens outlooks for the search for efficient terahertz rectifiers among Weyl semimetals and other topological materials with diverging Berry curvature.

In conclusion, we investigate the current-induced orbital Kerr effect and terahertz current responsivity in nonmagnetic, noncentrosymmetric materials displaying nonlinear Hall effect. In this case, the Kerr rotation is proportional to the a.c. BCD, and is therefore directly connected to the nonlinear Hall effect. The terahertz current responsivity is itself equal to half the orbital Kerr efficiency. We examined the orbital Kerr effect in selected noncentrosymmetric materials, using both effective models and realistic materials based on {\em ab initio} simulations, and compared our results with available experimental data, pointing out that the orbital Kerr efficiency tends to be larger in semiconductors than in metals, irrespective of the magnitude of the material's spin-orbit coupling. This opens particularly interesting promises for the observation of this effect in light compounds. In particular, our simulations on a metallic superlattice suggest that interfacial engineering can be used to optimize the nonequilibrium orbital Kerr effect, the nonlinear Hall effect, and its ability to detect terahertz radiations. Along the same line, van der Waals heterostructures \cite{Geim2013,Liu2016b} that combine materials with very different electronic properties (e.g., semiconducting, metallic, semimetallic) could also provide an interesting platform for the study of these effects.

\begin{acknowledgments}
D.G.O. and A. M. acknowledge support from the Excellence Initiative of Aix-Marseille Universit\'e - A*Midex, a French "Investissements d'Avenir" program. A.P. was supported by the ANR ORION project, grant ANR-20-CE30-0022-01 of the French Agence Nationale de la Recherche.
\end{acknowledgments}
\bibliography{Biblio2023}
\end{document}